\documentstyle[preprint,aps]{revtex}
\begin{document}
    \setlength{\baselineskip}{2.6ex}

\title{Pion-Nucleon Physics and the Polarizabilities of the Nucleon}
\author{Thomas R. Hemmert\thanks{Talk given at MENU97; Seventh International
Symposium on Meson-Nucleon Physics and the Structure of the Nucleon; Vancouver
 July 1997}\\
{\em TRIUMF Theory Group, 4004 Wesbrook Mall, Vancouver, B.C. Canada V6T 2A3}}

\maketitle

\begin{abstract}
\setlength{\baselineskip}{2.6ex}
I present recent results regarding the influence of 
pion-nucleon and nucleon-resonance
physics on the polarizabilities of the nucleon as measured in
Compton scattering.
\end{abstract}

\setlength{\baselineskip}{2.6ex}

\section*{INTRODUCTION}

In this workshop on pion-nucleon physics I am going to talk about Compton 
scattering off the nucleon, which may seem to be a strange topic to be 
presented here. However, I want to convince you that pion-nucleon and 
also $\Delta$(1232) dynamics are essential for understanding the results from
low energy Compton scattering--by this I mean photon energies in the c.m.
frame of less than 100 MeV.
Let us start with a very simple picture of the nucleon: A point particle
with spin 1/2 and an anomalous magnetic moment $\kappa$. In the 1950s Powell 
has already calculated the cross section for this model
and it turns out it describes the experimental data quite well for 
photon energies up to 50 MeV.

If one increases the energy of 
the incoming photon beam one starts seeing deviations from the simple 
Powell predictions, as one is picking up sensitivity to the 
internal structure of
the nucleon. In the past few years we have learned that this {\em low 
energy structure} of the nucleon can be described very well in 
terms of virtual
pion excitations around an unresolved spin 1/2 nucleon, together
with some contributions from nucleon resonances. The most precise method
to calculate these structure effects 
to this date is heavy baryon chiral perturbation theory 
(HBChPT), for a recent review see \cite{review}. If you wish, you can say
that HBChPT is an effective field theory with systematic power counting
that allows for a precise calculation of the ``pion cloud'' of the nucleon.    
In this presentation I will briefly outline the low energy structure of the 
Compton amplitude, define the polarizabilities of the nucleon and present
new predictions obtained in a ChPT framework with explicit pion, 
nucleon and delta degrees of freedom. For details and more references 
see \cite{prd,preprint}.

\section*{COMPTON SCATTERING AT LOW ENERGIES}

Assuming invariance under parity, charge conjugation and time reversal 
symmetry the general amplitude for Compton scattering off a proton $(\;\gamma
\;p\rightarrow \gamma^\prime \; p^\prime \;)$ can be written in terms 
of 6 structure dependent functions $A_i(\omega , \theta ), \; i=1..6$,
with $\omega = \omega^\prime$ denoting the photon energy and
$\theta$ being the scattering angle:
\begin{eqnarray}
T_{cms} &=& A_1(\omega,\theta)\;\vec{\epsilon}^{* \prime}\cdot\vec{\epsilon} 
+A_2(\omega,\theta)\;\vec{\epsilon}^{* \prime}\cdot\hat{k} \; \vec{\epsilon}
\cdot\hat{k}^\prime 
+A_3(\omega,\theta)\;i\vec{\sigma}\cdot(\vec{\epsilon}^{* \prime}\times
\vec{\epsilon}) \nonumber \\
& & +A_4(\omega,\theta)\;i\vec{\sigma}\cdot(\hat{k}^\prime \times\hat{k})
\vec{\epsilon}^{* \prime} \cdot\vec{\epsilon} 
+A_5(\omega,\theta)\;i\vec{\sigma}\cdot[(\vec{\epsilon}^{* \prime} \times
\hat{k}) \vec{\epsilon}\cdot\hat{k}^\prime -(\vec{\epsilon}\times
\hat{k}^\prime ) \vec{\epsilon}^{* \prime} \cdot\hat{k}]\nonumber\\
& & +A_6(\omega,\theta)\;i\vec{\sigma}\cdot[(\vec{\epsilon}^{* \prime}\times
\hat{k}^\prime ) \hat{\epsilon}\cdot\hat{k}^\prime -(\vec{\epsilon}\times
\hat{k})\vec{\epsilon}^{* \prime} \cdot\hat{k}]
\end{eqnarray}
Here $\vec{\epsilon},\hat{k}\; (\vec{\epsilon}^\prime ,\hat{k}^\prime )$ are 
the polarization vector, direction  of the incident (final) photon while 
$\vec{\sigma}$ represents the (spin) polarization vector of the nucleon.  
One now performs a low-energy expansion of the 6 independent  
functions $A_i(\omega,\theta)$ in powers of the photon energy $\omega$. For 
the case of a proton target of mass $M_N$ with anomalous magnetic moment 
$\kappa^{(p)}$ one finds  
\begin{eqnarray}
A_1(\omega,\theta)_{cms}&=&-{e^2\over M_N}+4\pi \left( \alpha_{E}^{(p)}+
\cos \theta \; \beta_{M}^{(p)}\right) \omega^2 
 -{e^2\over 4M_{N}^3}\left(1-\cos \theta\right) \omega^2 +
   \ldots \label{eq:a1} \\
A_2(\omega,\theta)_{cms}&=&{e^2\over M_{N}^2}\omega-4\pi\beta_{M}^{(p)}
\omega^2 +\ldots\\
A_3(\omega,\theta)_{cms}&=&\left[1+2\kappa^{(p)}-(1+\kappa^{(p)})^2\cos
\theta\right]{e^2\over 2M_{N}^2}\omega
-{(2\kappa^{(p)}+1)e^2\over 8M_{N}^4} \; \cos \theta \; \omega^3\nonumber\\
& &+ 4\pi \left[ \gamma_{1}^{(p)}-(\gamma_{2}^{(p)}+2\gamma_{4}^{(p)}) \cos 
\theta \right]\omega^3 +\ldots \label{eq:a3} \\
A_4(\omega,\theta)_{cms}&=&-{(1+\kappa^{(p)})^2e^2\over 2M_{N}^2}\omega+4\pi
\gamma_{2}^{(p)}\omega^3+\ldots\\
A_5(\omega,\theta)_{cms}&=&{(1+\kappa^{(p)})^2e^2\over 2M_{N}^2}\omega+4\pi
\gamma_{4}^{(p)}\omega^3+\ldots\\
A_6(\omega,\theta)_{cms}&=&-{(1+\kappa^{(p)})e^2\over 2M_{N}^2}\omega+4\pi
\gamma_{3}^{(p)}\omega^3 +\ldots \label{eq:a6}
\end{eqnarray}
The leading terms in the 6 structure functions are completely model-independent
and coincide with the old low energy theorems of current algebra. The ``real''
structure dependence beyond the anomalous magnetic moment starts at sub-leading 
order in the $\omega$ expansion and
the associated 6 polarizabilities $\alpha_E, \; \beta_M, \; \gamma_1, \; 
\gamma_2,\;
\gamma_3, \; \gamma_4$ cannot be determined by symmetry considerations. In
{\em unpolarized} Compton scattering the electric polarizability $\alpha_E$ 
and the magnetic polarizability $\beta_M$ describe the 
leading structure dependent effects and account for the deviation of the cross
section from the Powell result. The most recent fits yield \cite{ab}
\begin{eqnarray}
\alpha_E^{(p)}=(12.1\pm 0.8\pm 0.5)\times 10^{-4}\,{\rm fm}^3 ,
\quad \quad
\beta_M^{(p)}=(2.1\mp 0.8\mp 0.5)\times 10^{-4}\,{\rm fm}^3 ,
\label{eq:za}
\end{eqnarray}
indicating that the nucleon is a rather ``stiff'' object that cannot easily 
be deformed in the electric and magnetic field of the incoming and outgoing
photon.

While the linear response to external electric and magnetic fields 
$(\vec{E}, \vec{B})$ for a
classical (macroscopic) object is uniquely determined by the 2 
polarizabilities $\alpha_E,
\beta_M$, the nucleon due to its extra spin 1/2 degree of freedom has
4 additional response parameters $\gamma_i$ in external $\vec{E}, \vec{B}$ 
fields, commonly called the ``spin-polarizabilities'' of the nucleon 
\cite{ragusa}. All 6 structure dependent parameters are intrinsic properties
of the nucleon and their determination in Compton scattering therefore amounts
to a test of (low energy) QCD.

There exists a long history of experiments trying to determine 
$\alpha_E,\; \beta_M$, whereas the 4 spin-polarizabilities have only recently 
attracted the attention of experimentalists, as one requires polarized photon
sources in addition to polarized targets and has to measure over a wide range
of scattering angles $\theta$ in order to extract the $\gamma_i$ contributions.
In the absence of double-polarization experiments one has nevertheless tried
to obtain some estimates of particular linear combinations of the 4 
$\gamma_i$ from
multipole analyses in the single pion production region and unpolarized 
Compton scattering in the backward direction. This is not the place to comment
in detail on these ``experimental'' determinations, but I want to express
a strong caveat that the quoted errors could be severely underestimated due to 
strong model-dependencies of the extraction process. Keeping this in mind,
the current knowledge of spin-polarizabilities from {\em unpolarized} data 
reads \cite{SWK}
\begin{eqnarray}
\gamma_0^{(p)}&=&\gamma_1-\gamma_2-2\gamma_4\approx - 1.34 \times 10^{-4} \; 
                 {\rm fm}^4 \; , \label{eq:exp} \\
\gamma_\pi^{(p)}&=&\gamma_1+\gamma_2+2\gamma_4 \approx - \left( 28.0\pm2.8
                   \pm2.5\right) \times 10^{-4} \; {\rm fm}^4 \; .
\end{eqnarray}     
Further details on the current experimental situation can be found in 
\cite{preprint}. The huge numerical difference between $\gamma_0$ and 
$\gamma_\pi$ can be understood if one analyses the underlying physics using
ChPT.

\section*{THE PHYSICS BEHIND THE POLARIZABILITIES}
 
In 1992 it was found in a ${\cal O}(p^3)$ HBChPT calculation \cite{structure}
that ChPT can very nicely explain the magnitude of both $\alpha_E$ and 
$\beta_M$ as being dominated by $\pi N$ loop effects. According to this 
interpretation the only structure of the nucleon a low energy photon resolves
when undergoing Compton scattering would therefore be given by the nucleon's 
``pion-cloud'', in marked contrast to analyses using dispersion 
relations \cite{lvov}. A subsequent ${\cal O}(p^4)$ calculation \cite{BKSM}
proved that there are indeed only 
small corrections to the ${\cal O}(p^3)$ result, yielding
\begin{eqnarray}
\alpha_{E}^{(p)}|_{O(p^4)}=\left(10.5\pm2.0\right) \times 10^{-4} \; 
                             {\rm fm}^3 , \quad \quad
\beta_{M}^{(p)}|_{O(p^4)}=\left(3.5 \pm 3.6\right) \times 10^{-4} \;
                             {\rm fm}^3 . 
\end{eqnarray}
We note the agreement with current experimental results 
(Eq.{\ref{eq:za}), but also that there exists a considerable 
theoretical uncertainty. The main reason for this uncertainty is the first 
nucleon resonance $\Delta$(1232), which in HBChPT can only be
included via counterterms, i.e. is taken to be infinitely heavy compared to
the nucleon.
Recently a systematic formalism has been developed to include the delta as
an explicit degree of freedom in ChPT, called the ``small scale expansion''
\cite{letter}.
Herein one organizes the calculation in powers of $\epsilon$, which denotes
either a soft momentum, the pion mass or the nucleon-delta mass splitting.
The 6 polarizabilities of the nucleon have been calculated to ${\cal O}
(\epsilon^3)$ within this approach, taking into account all contributions 
arising from $\pi N$ loops, $\Delta$-pole graphs, $\pi\Delta$ loops and
neutral pion exchange via the anomalous $\pi^0\gamma\gamma$ vertex. The
pertinent Feynman diagrams and a discussion of the technical aspects regarding
calculations in this formalism can be found in \cite{prd}. Using
a new determination of the relevant $N\Delta$ coupling parameters
one finds the spin-independent polarizabilities
\begin{eqnarray}
\alpha_{E}^{(p)}|_{O(\epsilon^3)}=\left[ 12.2({\rm N\pi-loop})+0
                                        (\Delta-{\rm pole}) +4.2(\Delta\pi-
                                        {\rm loop})+0({\rm anom.})\right]
                                        \times 10^{-4}\,{\rm fm}^3 \; , 
                                  \label{eq:ae}\\
\beta_{M}^{(p)}|_{O(\epsilon^3)}=\left[1.2({\rm N\pi-loop})+7.2(\Delta-
                                    {\rm pole})+0.7(\Delta\pi-{\rm loop})
                                    +0({\rm anom.})\right]\times 10^{-4}\,
                                    {\rm fm}^3 \; . \label{eq:bm}
\end{eqnarray}
A quick glance at these results shows that the ${\cal O}(\epsilon^3)$ 
calculation is not able to reproduce the experimental results. In particular,
the large diamagnetic ``recoil'' contribution of the $\pi N$ loops in the
case of $\beta_M$ is only entering at ${\cal O}(p^4)$ in HBChPT 
\cite{BKSM} and thus necessitates a ${\cal O}(\epsilon^4)$ calculation for
a cancelation of the large paramagnetism of the $\Delta$-pole contribution
in Eq.\ref{eq:bm}.
Though numerically discouraging at this order, the solution to this old problem
in calculations of $\beta_M$ is thus known from the ${\cal O}(p^4)$ calculation
and is expected to work as well at ${\cal O}(\epsilon^4)$ in the small scale
expansion. The more ``troubling'' aspect of Eqs.\ref{eq:ae}f is
actually the large contribution from the $\pi\Delta$ continuum to $\alpha_E$.
In HBChPT it is very common to subsume pole contributions from nucleon 
resonances in counterterms, but there is no agreement in the chiral community
yet how one would include $\pi N^\ast$ or $\pi \Delta^{(*)}$ loop effects
in counterterms at a given order. Usually these effects are quite small and
can be safely neglected, but
the ${\cal O}(\epsilon^3)$ calculation of $\alpha_E$ shows a strong 
counterexample. A future ${\cal O}(\epsilon^4)$ calculation will therefore
shed more light on
the underlying physics in $\alpha_E$ and the issue of resonance saturation
of counterterms in the baryon sector in general. 

I now move on to discuss the physics of the spin-polarizabilities. In 
HBChPT they had been calculated to ${\cal O}(p^3)$ \cite{review}, but it was
quickly realized that $\Delta$(1232) could give large corrections. Now, unlike
the case of $\alpha_E, \beta_M$ where $\Delta$-pole contributions could be
incorporated at ${\cal O}(p^4)$ via counterterms, in the case of the 
spin-polarizabilities one would have to go to ${\cal O}(p^5)$, i.e. 2-loop, to
saturate the counterterms with delta exchange in a {\em complete} calculation.
I am sure you know that 2-loop
calculations in the baryon sector are outside today's ChPT technology, so the
spin-polarizabilities were ``forgotten'' for a while except for some occasional
phenomenological modeling. With the advent of the ``small scale expansion''
method the situation finally changed--it is now possible to systematically
calculate the effects of $\Delta$(1232) on the $\gamma_i$ with 1-loop 
technology. I present here the results of a recent ${\cal O}(\epsilon^3)$
calculation \cite{preprint}:
\begin{eqnarray}
\gamma_{1}^{(p)}&=\left[ 4.6({\rm N\pi-loop})+0
                                        (\Delta-{\rm pole}) -0.2(\Delta\pi-
                                        {\rm loop})-22({\rm anom.})\right]
                                        \times 10^{-4}\,{\rm fm}^4  
                                        \label{eq:g1} \\
\gamma_{2}^{(p)}&=\left[ 2.3({\rm N\pi-loop})-2.4
                                        (\Delta-{\rm pole}) -0.2(\Delta\pi-
                                        {\rm loop})+0({\rm anom.})\right]
                                        \times 10^{-4}\,{\rm fm}^4  \\
\gamma_{3}^{(p)}&=\left[ 1.2({\rm N\pi-loop})+0
                                        (\Delta-{\rm pole}) -0.1(\Delta\pi-
                                        {\rm loop})+11({\rm anom.})\right]
                                        \times 10^{-4}\,{\rm fm}^4 \\
\gamma_{4}^{(p)}&=\left[ -1.2({\rm N\pi-loop})+2.4
                                        (\Delta-{\rm pole}) +0.1(\Delta\pi-
                                        {\rm loop})-11({\rm anom.})\right]
                                        \times 10^{-4}\,{\rm fm}^4 
\end{eqnarray}
Note that 3 of the 4 polarizabilities are dominated by neutral pion exchange
coupled with the anomalous $\pi^0\gamma\gamma$ vertex. In addition to this 
well-understood contribution there are strong interference effects between
$\pi N$ loops and $\Delta$(1232) pole graphs, whereas the $\pi\Delta$ 
continuum shows very little influence in the spin-sector. Comparing with 
the known ``experimental'' determinations Eqs.\ref{eq:exp}f one finds
\begin{eqnarray}
\gamma_{0}^{(p)}|_{O(\epsilon^3)}=+2.0 \times 10^{-4}\,{\rm fm}^4 ,
\quad \quad
\gamma_{\pi}^{(p)}|_{O(\epsilon^3)}=-37.2 \times 10^{-4}\,{\rm fm}^4 ,
\end{eqnarray}
which reproduces the dramatic difference in size between these 2 linear
combinations of spin-polarizabilities (anomaly contributions cancel 
{\em exactly} in $\gamma_0$ and are {\em maximal} in $\gamma_\pi$), but are
not in very good numerical agreement. As mentioned before, the results 
Eqs.\ref{eq:exp}f were extracted from {\em unpolarized} experiments and 
should be checked in a planned \cite{mainz} double-polarization experiment, 
whereas on the theoretical
side one has to study possible ${\cal O}(\epsilon^4)$ corrections to 
Eqs.\ref{eq:g1}ff to judge the convergence of the perturbation series.

\section*{CONCLUSION}

I have presented recent results for the 6 polarizabilities of the proton
calculated to ${\cal O}(\epsilon^3)$ in the ``small scale expansion'' of 
ChPT which modify the simple picture of the polarizabilities as just being 
a $\pi N$ loop effect in HBChPT.
${\cal O}(\epsilon^4)$ calculations are 
called for to get a better understanding of the underlying physics 
, whereas on the experimental side a new experimental program
has to start in order to determine the poorly known spin-polarizabilities. 
  
\section*{ACKNOWLEDGEMENTS}

I thank the organizers of MENU97 for 
the opportunity to present this work to the $\pi N$ community. Many thanks
also go to my collaborators Barry Holstein, Joachim Kambor and 
Germar Kn{\"o}chlein and to my colleagues in the TRIUMF Theory Group.

\bibliographystyle{unsrt}

\begin{thebibliography}{99}
\bibitem{review} V. Bernard, N. Kaiser, and U.-G. Mei\ss ner, ``Chiral 
Dynamics in Nucleons and Nuclei,'' Int. J. Mod. Phys. {\bf E4}, 193 (1995).
\bibitem{prd} T.R. Hemmert, B.R. Holstein, and J. Kambor, ``$\Delta$(1232)
and the polarizabilities of the nucleon,'' Phys. Rev. {\bf D55}, 5598 (1997).
\bibitem{preprint} T.R. Hemmert et al., 
``Compton Scattering and the Spin Structure of the Nucleon at 
Low Energies,'' TRIUMF preprint TRI-PP-97-20.
\bibitem{ab} {\em e.g.} 
B.E. MacGibbon et al., ``Measurement of the electric and magnetic
polarizabilities of the proton'' Phys. Rev. {\bf C52}, 2097 (1995). 
\bibitem{ragusa} S. Ragusa, ``Third-order spin polarizabilities of the 
nucleon,'' Phys. Rev. {\bf D47}, 3757 (1993).
\bibitem{SWK} A.M. Sandorfi, C.S. Whisnant, and M. Khandaker, 
``Incompatibility of multipole predictions for the nucleon spin-polarizability
and Drell-Hearn-Gerasimov sum rules,'' Phys. Rev. {\bf D50}, R6681 (1994);
and Talk given at 1997 APS meeting.
\bibitem{structure} V. Bernard et al., ``Chiral structure of the nucleon,''
Nucl. Phys. {\bf B388}, 315 (1992).
\bibitem{lvov} {\em e.g.} 
A.I. L'vov, ``Theoretical aspects of the polarizability of the
nucleon,'' Int. J. Mod. Phys. {\bf A8}, 5267 (1993). 
\bibitem{BKSM} V. Bernard et al., ``Aspects of nucleon Compton scattering,''
Z. Phys. {\bf A348}, 317 (1994).
\bibitem{letter} T.R. Hemmert, B.R. Holstein, and J. Kambor, ``Systematic 1/M
expansion for spin 3/2 particles in baryon ChPT,'' Phys. Lett. {\bf B395}, 89
(1997).
\bibitem{mainz} R. Miskimen, and M. Pavan, private communication.
\end{thebibliography}

\end{document}